\documentclass[12pt]{article}

\usepackage{array} 
\usepackage{amssymb}
\usepackage{graphics,graphpap}
\usepackage{graphicx}
\usepackage{color}
\usepackage{graphicx}
\usepackage{dcolumn}
\usepackage{epsfig}
\usepackage{bm}
\usepackage{amsmath}
\usepackage{amsfonts}
\usepackage{textcomp}
\usepackage{setspace}

\newcommand{\beq}{\begin{eqnarray}}
\newcommand{\eeq}{\end{eqnarray}}

\newcommand{\bmp}{\noindent\begin{minipage}{16cm}}
\newcommand{\emp}{\end{minipage}\vskip 7mm} 

\def\drawbox#1#2{\hrule height#2pt
        \hbox{\vrule width#2pt height#1pt \kern#1pt
              \vrule width#2pt}
              \hrule height#2pt}

\def\Asym#1#2{\vcenter{\vbox{\drawbox{#1}{#2}
              \kern-#2pt 
              \drawbox{#1}{#2}}}}

\def\simge{\mathrel{%
   \rlap{\raise 0.511ex \hbox{$>$}}{\lower 0.511ex \hbox{$\sim$}}}}

\def\simle{\mathrel{
   \rlap{\raise 0.511ex \hbox{$<$}}{\lower 0.511ex \hbox{$\sim$}}}}

\def\s#1{\setbox0=\hbox{$#1$}%
\rlap{\ifdim\wd0>.7em\kern.22\wd0\else\kern.1\wd0\fi /}#1}

\newcommand{\Slash}[1]{#1\!\!\!/}

\begin{document}

\begin{titlepage}
\title{\vspace*{-2.0cm}
\bf\Large
Continuum photon spectrum from $\boldsymbol{Z^1 Z^1}$ annihilations in universal extra dimensions
\\[5mm]\ }

\author{
Henrik Melb\'eus\thanks{email: \tt melbeus@kth.se},~~~
Alexander Merle\thanks{email: \tt amerle@kth.se},~~~and~~
Tommy Ohlsson\thanks{email: \tt tommy@theophys.kth.se}\\ \\
{\normalsize \it Department of Theoretical Physics, School of Engineering Sciences,}\\
{\normalsize \it Royal Institute of Technology (KTH) -- AlbaNova University Center,}\\
{\normalsize \it Roslagstullsbacken 21, 106 91 Stockholm, Sweden}\\
}
\date{\today}
\maketitle
\thispagestyle{empty}

\begin{abstract}
\noindent
We calculate the continuum photon spectrum from the pair annihilation of a $Z^1$ LKP in non-minimal universal extra dimensions. We find that, due to the preferred annihilation into $W^+ W^-$ pairs, the continuum flux of collinear photons is relatively small compared to the standard case of the $B^1$ as the LKP. This conclusion applies in particular to the spectral endpoint, where also the additional fermionic contributions are not large enough to increase the flux significantly. When searching for the line signal originating from $Z^1 Z^1$ annihilations, this is actually a perfect situation, since the continuum signal can be regarded as background to the smoking gun signature of a peak in the photon flux at an energy that is nearly equal to the mass of the dark matter particle. This signal, in combination with (probably) a non-observation of the continuum signal at lower photon energies, constitutes a perfect handle to probe the hypothesis of the $Z^1$ LKP being the dominant component of the dark matter observed in the Universe.
\end{abstract}

\end{titlepage}

\section{\label{sec:intro}Introduction}

Since Zwicky's first idea of dark matter (DM) in 1933~\cite{Zwicky:1933gu}, this field has advanced tremendeously on theoretical and observational grounds, so that it is now a fully accepted fact that most of the matter in the Universe must actually be dark~\cite{Komatsu:2010fb}. It is, however, still under debate what this mysterious DM indeed consists of, and not even its mass is known to a large extent. One of the most popular classes of DM is non-relativistic (cold) dark matter (CDM) consisting of weakly interacting massive particles (WIMPs) with a mass in the GeV to TeV range, i.e., particles that are charged under $SU(2)_{\rm L}$, or have at least a comparable interaction strength. Probably the most generic and definitely the most intensely studied candidate particle for a WIMP is the neutralino in supersymmetric extensions~\cite{Ellis:2010kf} of the Standard Model (SM).

However, there are also other theories that can yield WIMP candidate particles. Perhaps the most interesting alternatives to supersymmetric theories are theories with additional spatial dimensions, often referred to as Kaluza--Klein (KK) theories. A particularly simple such theory is the model of universal extra dimensions (UEDs)~\cite{Appelquist:2000nn}. Similar to the situation in supersymmetry, these theories stabilize the lightest KK particle (LKP) through the conservation of a new parity-like quantum number (KK-parity), which would render any electrically neutral LKP a good DM candidate. In the minimal UED model (MUEDs), the LKP turns out to be the first KK-excitation $B^1$ of the $U(1)_{\rm Y}$ gauge boson~\cite{Cheng:2002iz}, and this particle is indeed a potential CDM candidate~\cite{Servant:2002aq,Cheng:2002ej}: In order to obtain a value of its relic abundance that is in agreement with current observational constraints~\cite{Komatsu:2010fb}, its mass should be between 500~GeV and 1600~GeV~\cite{Burnell:2005hm,Kong:2005hn}, where the inclusion of higher KK-modes tends to slightly pull this range to higher masses~\cite{Belanger:2010yx}.

Turning the attention beyond MUEDs, one could also introduce non-trivial boundary localized terms that would enable other particles than the $B^1$ to be the LKP~\cite{Flacke:2008ne}. Among those new possibilities is the first KK-excitation of the neutral component of the $SU(2)_{\rm L}$ gauge field~\cite{Arrenberg:2008wy,Flacke:2009eu,Blennow:2009ag,Bonnevier:2011km}, which is usually denoted $Z^1$. It is this candidate that we are going to investigate in the present work. The most important investigation of this particle was the determination of its relic abundance~\cite{Arrenberg:2008wy}, which translates into an allowed mass range of roughly 1800~GeV to 2700~GeV. Furthermore, indirect annihilation signals of this DM candidate have been investigated, including annihilation into neutrinos~\cite{Blennow:2009ag} or into pairs of photons~\cite{Bonnevier:2011km}, the latter resulting into a monoenergetic peak. This line signal is of particular interest, since there are several experiments on the way aiming to detect the corresponding gamma-ray signals (e.g., Fermi-LAT~\cite{Michelson:2010zz}, H.E.S.S.~\cite{Abramowski:2011hc}, MAGIC~\cite{Aleksic:2011jx}, VERITAS~\cite{Acciari:2010pja}, CANGOROO-III~\cite{Kabuki:2007am}).

Current experimental bounds on DM consisting of a $Z^1$ LKP are relatively weak. The most recent and also the strongest direct detection limit is provided by the XENON100 experiment~\cite{Aprile:2011hi}, and it constrains the spin-independent DM-nucleon cross section. This quantity has already been calculated for $Z^1$ DM in Ref.~\cite{Arrenberg:2008wy}. Note that the parameter values used there are similar to the ones used in this letter. Comparing these results to the limit obtained from XENON100, we find that, for a relative mass splitting between the $Z^1$ and the first-level KK quarks larger than a few percent, the model would constrained for $M_{Z^1}$ below about 1~TeV. This value, however, is far too small to yield the correct relic abundance. Furthermore, in Ref.~\cite{Blennow:2009ag}, it has been found that the indirect neutrino signal from annihilations of $Z^1$ DM particles in the Sun is too weak to be observable in current neutrino telescopes. On the other hand, the indirect photon signal looks much more promising~\cite{Bonnevier:2011km}.

There is, however, an important ingredient that has not been calculated yet: Although Ref.~\cite{Bonnevier:2011km} has treated the peaked line signal for $Z^1 Z^1$ annihilations, a calculation of the continuum photon spectrum has, to our knowledge, not been performed before for the $Z^1$. The continuum spectrum can arise from photons coupling to electrically charged final or intermediate states in annihilation processes of $Z^1 Z^1$ pairs into two SM particles, which are the only channels allowed by KK-number conservation. Apart from constituting a signal by itself, this continuum spectrum can also be viewed as ``background'' to the peak spectrum since it is, due the finite energy resolution, experimentally not necessarily possible to resolve the peak above the continuum. In such a case, a continuum flux that is too high could destroy the chance to use the peak flux in order to directly extract information about the $Z^1$ mass. In this manuscript, we will close the remaining gap by presenting a calculation of the continuum spectrum arising from mostly collinear photons.

This paper is organized as follows: In Sec.~\ref{sec:calc}, we describe how to obtain the continuum spectrum, and in Sec.~\ref{sec:res}, we present our numerical results. Finally, in Sec.~\ref{sec:conc}, we draw our conclusions.

\section{\label{sec:calc}Final state radiation in $\boldsymbol{Z^1 Z^1}$ annihilations}

Final state radiation (FSR) in WIMP-WIMP annihilations can arise if the annihilation process contains electrically charged SM final (or intermediate) states $X$, to which a photon can couple. The first crucial point is that, due to the photon being massless, the emission of photons will always be possible whenever annihilation into a pair $X \overline{X}$ is kinematically possible. Next, since the mass of the WIMP is usually much larger than the mass of any SM particle, the final state particles will in general be highly relativistic, which causes the final state photons to be collinear with either $X$ or $\overline{X}$, to a very good approximation. An excellent treatment of these matters can be found in Ref.~\cite{Birkedal:2005ep}, which we will follow closely. We will apply the methods introduced in that paper to analyze the situation for $Z^1 Z^1$ annihilations in non-minimal UEDs.

The decisive point is that the annihilation cross section including FSR factorizes in the following way:
\begin{equation}
 \frac{{\rm d} \sigma(Z^1 Z^1\to X \overline{X} \gamma)}{{\rm d}x} \approx \frac{\alpha Q_X^2}{\pi} \mathcal{F}_X(x) \log \left[ \frac{s (1-x)}{m_X^2} \right]\ \sigma(Z^1 Z^1\to X \overline{X}),
 \label{eq:FSR_1}
\end{equation}
where $x=2 E_\gamma/\sqrt{s}=E_\gamma/M_{Z^1}$, with $E_\gamma$ being the photon energy and $s$ the center-of-mass energy squared. The fine-structure constant $\alpha$ should, in principle, be run up to the TeV scale, but since an energy of a few TeV is just one order of magnitude larger than the energy at the $Z$-pole, it is enough to use the corresponding value of $\alpha\approx 1/128$. Since the $Z^1$ is assumed to be the LKP (it could not be the dark matter particle otherwise, since it would be unstable) and since KK-number conservation forces the final states to be even under KK-parity, we can have tree-level annihilations only into $X \overline{X}$ pairs, which are contained in the list $\{ e\overline{e}, \mu\overline{\mu}, \tau\overline{\tau}, u\overline{u}, d\overline{d}, c\overline{c}, s\overline{s}, t\overline{t}, b\overline{b}, W^+ W^-, \nu_e \overline{\nu}_e, \nu_\mu \overline{\nu}_\mu, \nu_\tau \overline{\nu}_\tau, Z^0 Z^0, H H^* \}$. Of course, although $Z^1 Z^1$ pairs can annihilate into neutrinos~\cite{Blennow:2009ag} or other electrically neutral particles, these processes will not contribute to Eq.~\eqref{eq:FSR_1}, since $Q=0$. The splitting function $\mathcal{F}_X(x)$ is given by~\cite{Birkedal:2005ep}
\begin{equation}
 \begin{matrix}
 \mathcal{F}_F(x)=\frac{1+(1-x)^2}{x} & {\rm for\ fermions,} \\
 \mathcal{F}_B(x)=\frac{1-x}{x}\hfill \hfill & {\rm for\ bosons.}\hfill \hfill
 \end{matrix}
 \label{eq:FSR_2}
\end{equation}
Note that, due to the final states being highly relativistic, vector final states will practically act as scalars (and hence have the same splitting function), which is a reflection of the well-known Goldstone boson equivalence theorem~\cite{Cornwall:1974km,Vayonakis:1976vz,Lee:1977eg}.

The quantity we are actually interested in is the differential photon multiplicity~\cite{Bergstrom:2004cy,Bringmann:2007nk} for each final state,
\begin{equation}
 \frac{{\rm d} N_\gamma^{X\overline{X}}}{{\rm d}x} = \frac{{\rm d} \sigma(Z^1 Z^1\to X\overline{X} \gamma)/{\rm d}x}{\sigma(Z^1 Z^1\to X \overline{X})}.
 \label{eq:FSR_3}
\end{equation}
Due to the factorization in Eq.~\eqref{eq:FSR_1}, the 2-body annihilation cross section drops out of this quantity, which essentially means that the shape of the spectrum does not depend on the actual rate. However, to calculate the exact value of the spectrum we do need the cross sections.

In order to obtain the actual flux, we have to calculate~\cite{Bergstrom:2004cy,Bergstrom:1997fj}
\begin{equation}
 \frac{{\rm d} \Phi_\gamma}{{\rm d} E_\gamma} \simeq \frac{3.5\cdot 10^{-8}}{M_{Z^1}^2} \frac{{\rm d} N_\gamma^{\rm eff}}{{\rm d}x} \left( \frac{\sigma_{\rm tot} v_{\rm rel}}{3\cdot 10^{26}\ {\rm cm}^{-3} {\rm s}^{-1}} \right) \left( \frac{0.8~{\rm TeV}}{M_{Z^1}} \right) \langle J_{\rm GC} \rangle_{\Delta \Omega} \Delta \Omega\ {\rm m^{-2} s^{-1} TeV^{-1}}.
 \label{eq:FSR_4}
\end{equation}
Here, the total number of photons per $Z^1 Z^1$ annihilation is given by
\begin{eqnarray}
 \frac{{\rm d} N_\gamma^{\rm eff}}{{\rm d}x} &\equiv& \sum_F \kappa_F \frac{{\rm d} N_\gamma^{F\overline{F}}}{{\rm d}x} + \sum_B \kappa_B \frac{{\rm d} N_\gamma^{B\overline{B}}}{{\rm d}x}\nonumber\\
 &\approx& \frac{\alpha}{\pi} \kappa_W \mathcal{F}_B(x) \log \left[ \frac{s (1-x)}{M_W^2} \right] + \sum_{F=l,q} \frac{\alpha Q_F^2}{\pi} \kappa_F \mathcal{F}_F(x) \log \left[ \frac{s (1-x)}{m_F^2} \right],
 \label{eq:FSR_5}
\end{eqnarray}
where the last sum runs over all electrically charged leptons and over all quarks. The quantity $\kappa_X$ denotes the branching ratio into $X \overline{X}$. To obtain expressions for the branching ratios, one can make use of the total cross section formulas given in the literature~\cite{Burnell:2005hm,Kong:2005hn,Hooper:2007qk}\footnote{Note that there are some typos in the expressions found in Ref.~\cite{Hooper:2007qk}, which can, however, easily be corrected when in addition consulting Ref.~\cite{Kong:2005hn}.} to calculate
\begin{equation}
 \kappa_X = \frac{\sigma (Z^1 Z^1\to X \overline{X})}{\sigma_{\rm tot}},
 \label{eq:FSR_6}
\end{equation}
where the total annihilation cross section is given by
\begin{eqnarray}
 \sigma_{\rm tot} &=& \sigma (Z^1 Z^1\to e^- e^+) + \sigma (Z^1 Z^1\to \mu^- \mu^+) + \sigma (Z^1 Z^1\to \tau^- \tau^+) \nonumber\\
 &+& \sigma (Z^1 Z^1\to \nu_e \overline{\nu}_e) + \sigma (Z^1 Z^1\to \nu_\mu \overline{\nu}_\mu) + \sigma (Z^1 Z^1\to \nu_\tau \overline{\nu}_\tau) \nonumber\\
 &+& \sigma (Z^1 Z^1\to u \overline{u}) + \sigma (Z^1 Z^1\to c \overline{c}) + \sigma (Z^1 Z^1\to t \overline{t})\nonumber\\
 &+& \sigma (Z^1 Z^1\to d \overline{d}) + \sigma (Z^1 Z^1\to s \overline{s}) + \sigma (Z^1 Z^1\to b \overline{b})\nonumber\\
 &+& \sigma (Z^1 Z^1\to H H) + \sigma (Z^1 Z^1\to W^- W^+) + \sigma (Z^1 Z^1\to Z^0 Z^0).
 \label{eq:FSR_7}
\end{eqnarray}
Note that there is no tree-level annihilation into $\gamma \gamma$, since the $Z^1$ is, naturally, not electrically charged. Analogously, there is also no tree-level annihilation into two gluons. In the actual computation, it is perfectly sufficient to calculate the expansion of $\sigma_{\rm tot}$ in terms of the relative velocity $v_{\rm rel}$ of the (non-relativistic) initial state WIMPs and use the lowest-order terms only.

To proceed, let us note that $\langle J_{\rm GC} \rangle_{\Delta \Omega} \Delta \Omega\simeq 0.13 b$ for $\Delta \Omega=10^{-5}$ in a Navarro--Frenk--White (NFW)~\cite{Navarro:1995iw} dark matter profile with parameters $(\alpha, \beta, \gamma, r_S)=(1.0, 3.0, 1.0, 20~{\rm kpc})$ in the galactic halo. The boost factor $b$ might enhance the signal for a profile that is clumpier than anticipated. However, we will stick to $b=1$ here.

\section{\label{sec:res}Numerical results}

\begin{figure}[t]
\centering
\includegraphics[width=10cm]{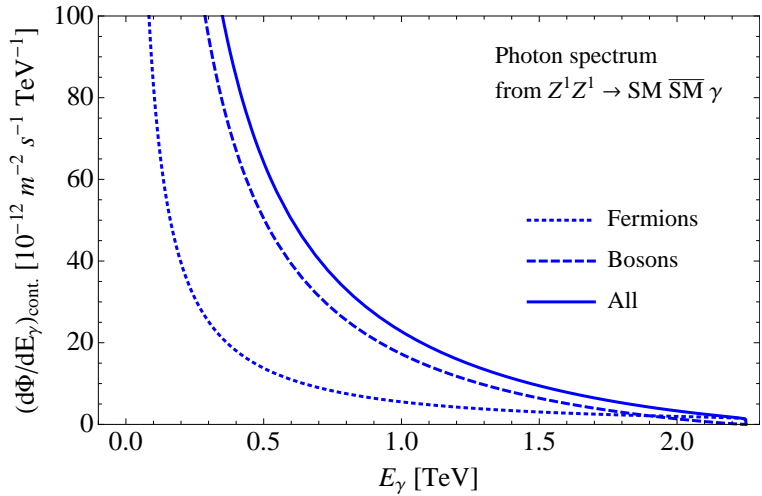}
\caption{\label{fig:Full} The full continuum photon spectrum for $Z^1 Z^1$ annihilations.}
\end{figure}

The result for the full spectrum is displayed in Fig.~\ref{fig:Full}, where we have assumed a $Z^1$ mass of $M_{Z^1}=2250$~GeV, which is just a suitable value in order to obtain the correct DM abundance~\cite{Arrenberg:2008wy,Bonnevier:2011km}. We have numerically checked that varying the $Z^1$ mass within the range allowed by the requirement of having the correct abundance does not qualitatively change our results. Note that all particle masses as well as the gauge coupling $g$ are taken from Ref.~\cite{Nakamura:2010zzi}. At first, it might seem odd that over practically the whole spectrum the FSR contribution from decays into $W^+ W^-$ pairs dominates, although this is the only boson-antiboson pair that comes into play, whereas there are (including color charge) $3+3\cdot 6=21$ fermion pairs into which $Z^1 Z^1$ could annihilate while simultaneously radiating off photons. This is confirmed by the results of, e.g., Ref.~\cite{Blennow:2009ag}, which also obtains a branching ratio of roughly $90~\%$ into $W^+ W^-$. However, it is not too much of a surprise when taking into account that we need to have a parity violation for the annihilation process to occur: The two identical (non-relativistic) vector bosons in the initial state will always have a parity of $P=+1$, while a fermion-antifermion pair will have a parity of $P=-1$, which causes the corresponding transition to be suppressed.\footnote{The mathematical version of this argument is that the large momentum contribution of the internal fermion propagator is canceled by the two identical projection operators, $P_{\rm L} (\Slash{p}+m) P_{\rm L}=m P_{\rm L}$, whereas only the smaller contribution proportional to the mass remains.} An annihilation into $W^+ W^-$, on the other hand, is not suppressed by any such reason including the conservation of angular momentum. Note that this is one of the major differences in comparison to the annihilation of the ``standard'' LKP $B^1$~\cite{Bergstrom:2004cy}, which has a much weaker coupling to $W^+ W^-$ due to its Abelian nature. In addition, the requirement of the $Z^1$ LKP having to be heavier than a $B^1$ LKP reduces the flux considerably: The cross sections themselves are, in the non-relativistic limit, proportional to $1/m_{\rm LKP}^2$~\cite{Burnell:2005hm,Kong:2005hn,Hooper:2007qk}, and the flux in Eq.~\eqref{eq:FSR_4} suffers from an additional proportionality to $1/m_{\rm LKP}^3$, which means that the flux of an LKP with a mass twice as large as the one of the LKP in an alternative scenario experiences a strong reduction by a factor of $1/2^5\approx 0.03$. A third reason why the detection prospects for the continuum spectrum from $Z^1 Z^1$ annihilations are much worse than for $B^1 B^1$ is that the logarithmic enhancements from Eq.~\eqref{eq:FSR_1} are much stronger for small final state masses, and hence for most of the fermions in the SM.\footnote{This can be easily understood by glancing at the well-known example of the harmonic oscillator in quantum mechanics: The frequency turns out to be inversely proportional to the square root of the mass, which essentially means that it is harder to make a heavier particle oscillate and hence radiate off (or absorb) a photon.} Annihilation into the fermions is, however, suppressed for $Z^1 Z^1$, since the non-Abelian nature of the gauge bosons causes them to annihilate very efficiently into $W^+ W^-$, whose contribution to the photon spectrum is weak.

\begin{figure}[t]
\centering
\begin{tabular}{lr}
\includegraphics[width=8.0cm]{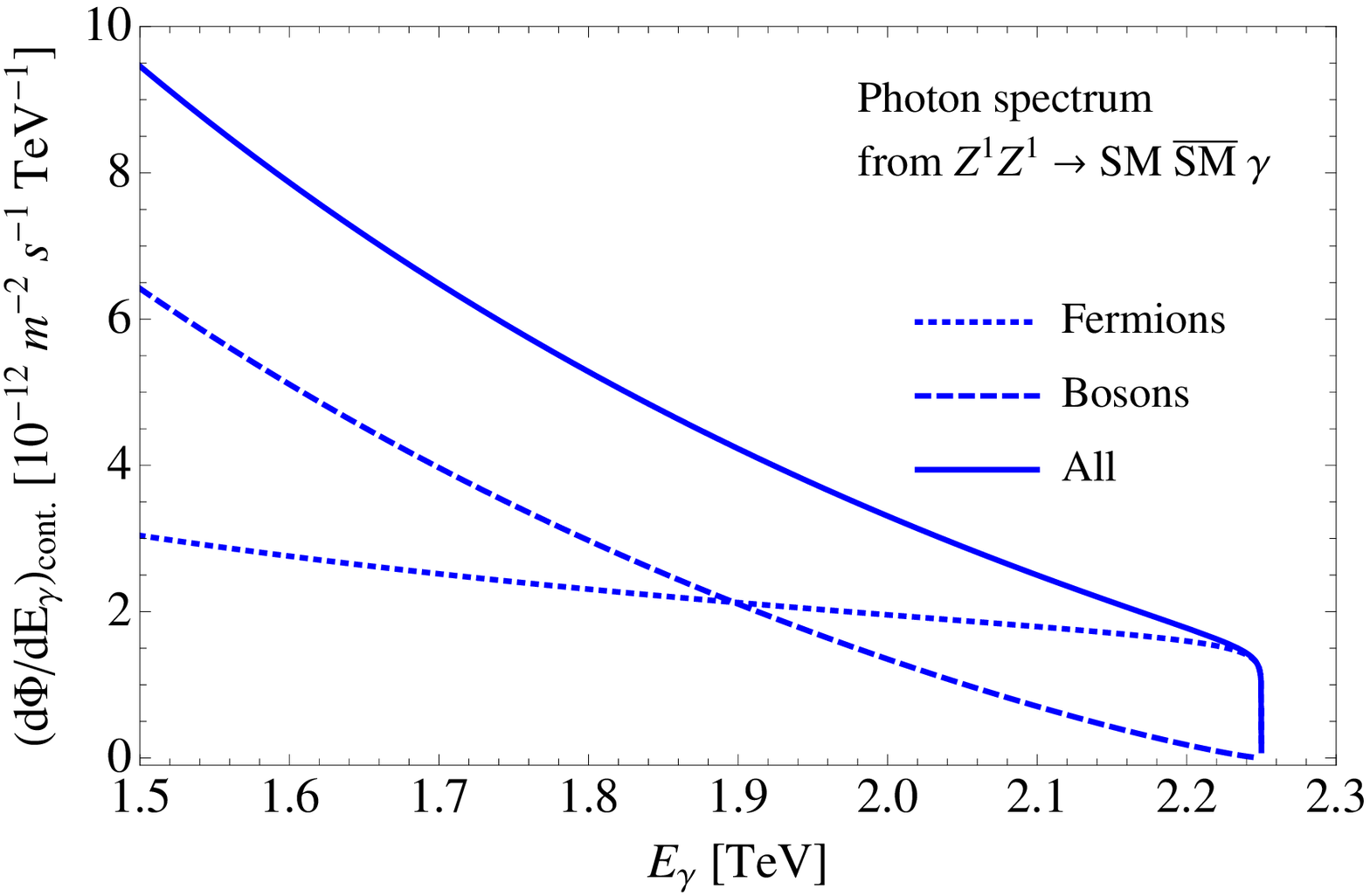} & \includegraphics[width=8.0cm]{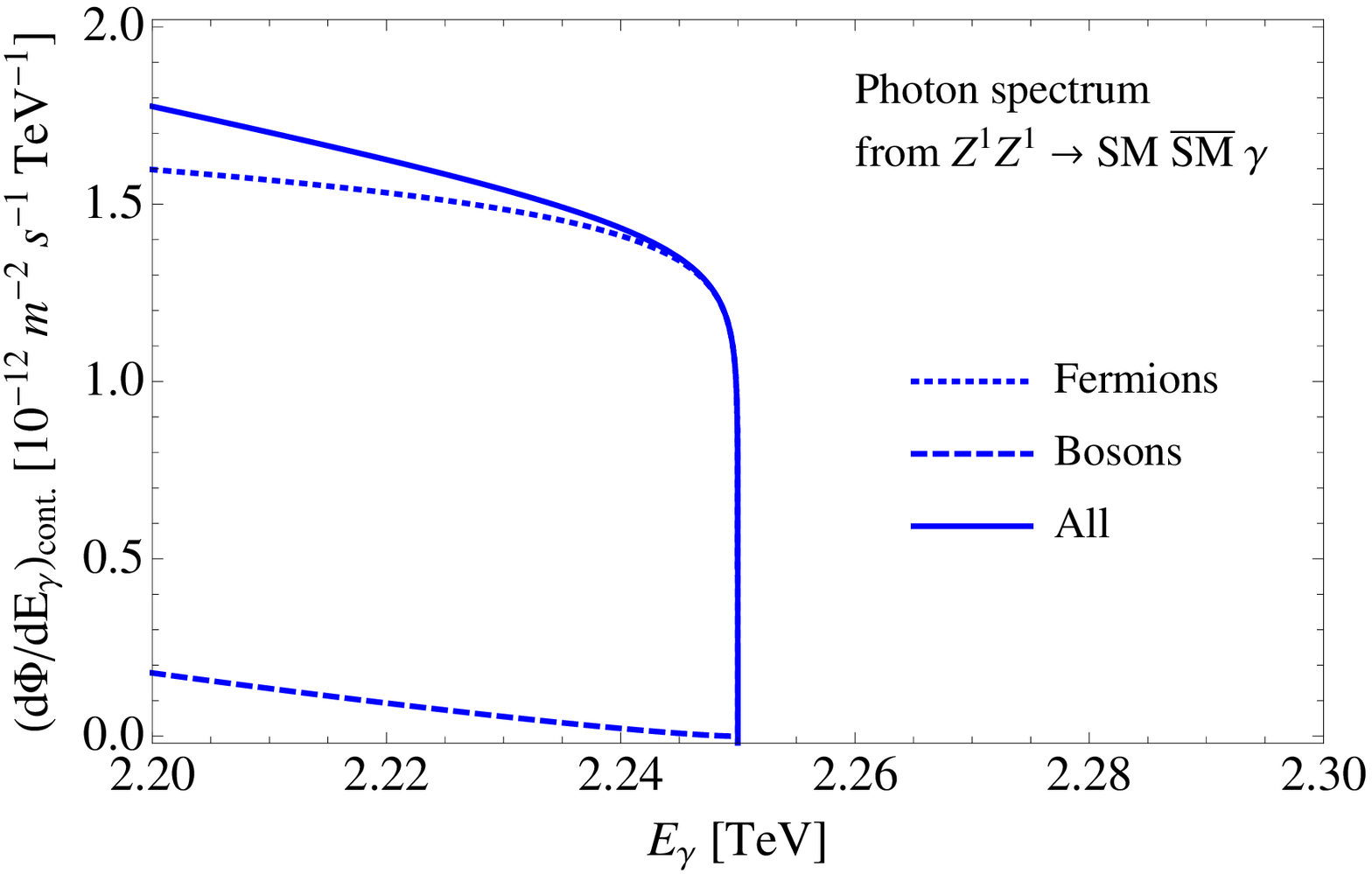}
\end{tabular}
\caption{\label{fig:Endpoint} The crossing and endpoint regions of the continuum photon spectrum for $Z^1 Z^1$ annihilations.}
\end{figure}

An interesting point to mention is the endpoint region, which is displayed in Fig.~\ref{fig:Endpoint}: As can be seen from the left panel, it turns out that close to the endpoint the situation is actually the opposite of the one described above. Indeed, at this end of the spectrum, the fermionic contribution dominates, although the branching ratios into fermions remain small. The reason for this is the different spectral shape of the fermion and boson splitting functions~\cite{Birkedal:2005ep}: While the spectrum of the fermionic part is dominated by the collinear contribution, the bosonic part is suppressed. Actually, there might be model-dependent non-collinear contributions, which we neglect here. However, even if these contributions did dominate the unsuppressed collinear contributions from the bosonic part by a factor of 100, which is a vast overestimation, they would not enhance the total result by more than one order of magnitude, which would still not change any conclusions about a possible detection of the peak signal. Also a rough estimate of this contribution at the endpoint as $\sigma (Z^1 Z^1 \to W^+ W^- \gamma) \sim \alpha\cdot \sigma (Z^1 Z^1 \to W^+ W^-)$, which is even too optimistic, since it neglects the additional phase space suppression of a 3-body final state as compared to a 2-body final state, results in a completely negligible perturbation to the case where this contribution is neglected. Furthermore, since the 2-body process is completely unsuppressed, we cannot expect any additional enhancement (as, e.g., from unlocking an angular momentum barrier) in the 3-body version with the photon included.
In terms of detection of a possible DM signature, this is actually a good sign~\cite{Bonnevier:2011km}: Although the kink arising from the collinear photons from annihilation into fermions (see right panel of Fig.~\ref{fig:Endpoint}) will not be as pronounced as for $B^1 B^1$ annihilations~\cite{Bergstrom:2004cy}, the line signal, which is the actual smoking gun signature of DM annihilations, will have excellent detection prospects~\cite{Bonnevier:2011km}. For the $Z^1$ mass considered, $M_{Z^1}=2250$~GeV, the peak signal will be stronger than the continuum signal by about four orders of magnitude. Note that, although the continuum spectrum would be enhanced for a lower $Z^1$ mass, cf.\ Eq.~\eqref{eq:FSR_4}, even the smallest possible values of $M_{Z^1}\approx 1800$~GeV still result in the peak being stronger by more than three orders of magnitude, which illustrates the robustness of our results.

Furthermore, the absence of photons at relatively low energies is a (negative) signal that can be correlated with the peak: The $Z^1$, in our setting, can be excluded as DM candidate if the peak is detected \emph{together} with the low-energy continuum spectrum. This, combined with the correct mass range derived from the relic density calculations~\cite{Arrenberg:2008wy}, offers a clear way to distinguish the $Z^1$ from the $B^1$ as a DM candidate.

\section{\label{sec:conc}Conclusions}

We have calculated the continuum photon spectrum for the $Z^1$ as the LKP in non-minimal UEDs. In addition to the photon and neutrino signals, the continuum spectrum is a third annihilation signal of major importance. We have shown that the continuum part of the signal is suppressed, due to the bad efficiency of the $W$-boson in contributing to the collinear photon spectrum. However, the $W$-bosons amount to roughly $90~\%$ of the annihilation products, which efficiently suppresses the flux of continuum photons. Close to the spectral endpoint, the contribution of the $W$-bosons is suppressed by the splitting function, causing the fermionic contribution to dominate in that region, but this contribution alone is not large enough to yield a strong signal.

Although our investigation shows that the detection prospects for the continuum signal itself are bad, this is actually good news when aiming to detect the peak signal from $Z^1 Z^1$ annihilations into two photons. Accordingly, one can hope to be able to either detect the peak signal in connection with the absence of the continuum signal in the near future, or one would have immediate prospects to rule out the hypothesis of the $Z^1$ being the LKP and constituting a major part of the DM in the Universe.

\section*{\label{sec:ack}Acknowledgements}

This work was supported by the Swedish Research Council (Vetenskapsr{\aa}det), contract no.\ 621-2008-4210 (T.O.), by the the G\"oran Gustafsson foundation (A.M.), and by the Royal Institute of Technology (KTH), project no.\ SII-56510.

\bibliographystyle{./apsrev}
\bibliography{./ContFlux}

\end{document}